\documentclass[prl,aps,twocolumn,showpacs,floatfix]{revtex4}

\usepackage{amssymb,graphicx}
\usepackage{epsfig}
\usepackage[usenames]{color}

\def\be{\begin{equation}}
\def\ee{\end{equation}  }
\def\bea{\begin{eqnarray}}
\def\eea{\end{eqnarray}  }

\begin{document}

\title{Black Strings, Low Viscosity Fluids, and Violation of Cosmic Censorship}

\author{Luis Lehner${}^{1,2,3}$ and Frans Pretorius${}^{4}$}
\affiliation{
${}^1$ Perimeter Institute for Theoretical Physics, Waterloo, Ontario N2L 2Y5, Canada\\
${}^2$ Department of Physics, University of Guelph, Guelph, Ontario N1G 2W1, Canada\\
${}^3$ Canadian Institute For Advanced Research (CIFAR), Cosmology and Gravity Program, Canada\\
${}^4$ Department of Physics, Princeton University, Princeton, NJ 08544,USA.
}


%
%
\begin{abstract}
We describe the behavior of 5-dimensional black strings, subject to the Gregory-Laflamme instability.
Beyond the linear level, the evolving strings exhibit a rich dynamics, where at intermediate stages
the horizon can be described as a sequence of 3-dimensional spherical black holes joined by black
string segments. These segments are themselves subject to a Gregory-Laflamme instability, resulting
in a self-similar cascade, where ever-smaller satellite black holes form connected by ever-thinner
string segments. This behavior is akin to satellite formation in low-viscosity fluid streams subject
to the Rayleigh-Plateau instability. The simulation results imply that the string segments will reach
zero radius in finite asymptotic time, whence the classical space-time terminates in a naked
singularity. Since no fine-tuning is required to excite the instability, this constitutes a generic
violation of cosmic censorship.
\end{abstract}


\maketitle

%
%
\noindent{\bf{\em Introduction:}}
While stationary black holes in 4 spacetime dimensions (4D) are 
 stable to perturbations, higher dimensional analogues are not. Indeed, as first illustrated by
Gregory and Laflamme in the early 90s~\cite{Gregory:1993vy}, black strings and 
p-branes are linearly unstable to long wavelength perturbations in 5 and
higher dimensions. Since then, a number of interesting black objects
in higher dimensional gravity have been discovered, many of them exhibiting 
similar instabilities (see e.g.~\cite{Emparan:2008eg}).

An open question for all unstable black objects is what the end-state of
the perturbed system is. For black strings~\cite{Gregory:1993vy} conjectured
that the instability would cause the horizon to pinch-off at periodic intervals, 
giving rise to a sequence of black holes. One reason for this conjecture comes from entropic
considerations: for a given mass per unit length and periodic spacing above a critical
wavelength $\lambda_c$, a sequence of hyperspherical black holes has higher entropy 
than the corresponding black string. Classically, event horizons cannot bifurcate
without the appearance of a naked singularity~\cite{Hawking:1973uf}. Thus, reaching the
conjectured end-state would constitute a violation of cosmic censorship, without 
``unnatural'' initial conditions or fine-tuning, and be an example
of a classical system evolving to a regime where quantum gravity is required.

This conjecture was essentially taken for granted until several years later when
it was proved that the generators of the horizon can
not pinch-off in finite affine time~\cite{Horowitz:2001cz}. From this,
it was conjectured that a new, non-uniform black string end-state would be reached~\cite{Horowitz:2001cz}. 
Subsequently, stationary, non-uniform black string
solutions were found~\cite{Gubser:2001ac,Wiseman:2002zc}, however, they had less entropy than the uniform string
and so could not be the putative new end-state,
at least for dimensions lower than 13~\cite{Sorkin:2004qq}.

A full numerical investigation studied 
the system beyond the linear regime~\cite{Choptuik:2003qd}, though not 
far enough to elucidate the end-state before the code ``crashed''.
At that point the horizon resembled
spherical black holes connected by black strings, though no
definitive trends could be extracted, still allowing for
both conjectured possibilities:
(a) a pinch-off in infinite affine time, 
(b) evolving to a new, non-uniform state.
If (a), a question arises whether pinch-off happens in
infinite {\em asymptotic} time; if so, any bifurcation
would never be seen by outside observers, and 
cosmic censorship would hold. While this might be a natural conclusion,
it was pointed out in~\cite{Garfinkle:2004em,Marolf:2005vn} that
due to the exponentially diverging rate between affine time and 
a well-behaved asymptotic time, pinch-off could
occur in finite asymptotic time.

A further body of (anecdotal) evidence supporting the GL conjecture
comes from the striking resemblance of the equations
governing black hole horizons to those describing fluid flows, the latter which
do exhibit instabilities that often result in break-up of the fluid. The fluid/horizon
connection harkens back to the 
membrane paradigm~\cite{Thorne:1986iy}, and also in
more recently developed correspondences~\cite{Bhattacharyya:2008jc,Emparan:2009at}.
In~\cite{Cardoso:2006ks} it was shown that the dispersion relation of Rayleigh-Plateau unstable
modes in hyper-cylindrical fluid flow with tension agreed well with those of the GL modes
of a black string.
Similar behavior was found for instabilities of a self-gravitating cylinder of
fluid in Newtonian gravity~\cite{Cardoso:2006sj}. 
In~\cite{Camps:2010br}, using a perturbative expansion of the Einstein field equations~\cite{Emparan:2009at}
to related the dynamics of the horizon to that of a viscous fluid,
the GL dispersion relation was {\em derived} to good approximation,
thus going one step further than showing analogous behavior between
fluids and horizons.

What is particularly intriguing about fluid analogies, and what they might
imply about the black string case,
is that break-up of an unstable flow is preceded by formation of spheres
separated by thin necks. 
For high viscosity liquids, a single neck forms before 
break-up. For lower viscosity fluids, smaller ``satellite'' spheres can form
in the necks, with more generations forming the lower the viscosity (see
~\cite{Eggers:1997zz} for a review). In the membrane paradigm, black holes have
lower shear viscosity to entropy ratio than any known fluid~\cite{Kovtun:2004de}.

Here we revisit the evolution of 5D black strings using a new code. 
This allows us to follow the evolution well beyond the earlier study~\cite{Choptuik:2003qd}. 
We find that the dynamics of the horizon
unfolds as predicted by the low viscosity fluid analogues: the string initially evolves
to a configuration resembling a hyperspherical black hole connected by thin
string segments; the string segments are themselves unstable, and the
pattern repeats in a self-similar manner to ever smaller scales. Due to finite
computational resources, we cannot follow the dynamics indefinitely. If
the self-similar cascade continues as suggested by the simulations, 
arbitrarily small length scales, and in consequence arbitrarily large curvatures
will be revealed outside the horizon in finite asymptotic time.

%
%
\noindent{\bf{\em Numerical approach:}} 
We solve the vacuum Einstein field equations
in a 5-dimensional (5D)
asymptotically flat spacetime with an $SO(3)$ symmetry. Since 
perturbations of 5D black strings violating this
symmetry are stable and decay~\cite{Gregory:1993vy}, we do not expect 
imposing this symmetry qualitatively affects the results
presented here. 

We use the generalized harmonic formulation of the field 
equations~\cite{Pretorius:2004jg}, and adopt a {\em Cartesian} coordinate 
system related to spherical polar coordinates via
 $\bar{x}^i=(\bar{t},\bar{x},\bar{y},\bar{z},\bar{w})=(t,r\cos\phi\sin\theta,r\sin\phi\sin\theta,r\cos\theta,z)$.
The black string horizon has topology $S^2\times R$; $(\theta,\phi$)
are coordinates on the 2-sphere, and $z$ ($\bar{w}$) is the coordinate
in the string direction, which we make periodic with length $L$. 
We impose a {\em Cartesian Harmonic} gauge condition,
i.e. $\nabla_\alpha \nabla^\alpha \bar{x}^i =0$,
as empirically this seems to result in more stable numerical
evolution compared to spherical harmonic coordinates.
The $SO(3)$ symmetry is enforced using the variant of the 
``cartoon'' method~\cite{Alcubierre:1999ab} described in~\cite{Pretorius:2004jg}, 
were we only evolve a $\bar{y}=\bar{z}=0$ slice of the spacetime.
We further add {\em constraint damping}~\cite{Gundlach:2005eh},
which introduces two parameters $\kappa$ and $\rho$; we use $(\kappa,\rho=1,-0.5)$,
where a non-zero $\rho$ is essential to damp an unstable zero-wavelength
mode arising in the $z$ direction. 

We discretize the equations using
4th order finite difference approximations, and integrate in
time using 4th order Runge-Kutta. 
To resolve the small length scales that develop during evolution
we use Berger and Oliger adaptive mesh refinement.
Truncation error estimates are used to dynamically generate the mesh hierarchy, and
we use a spatial and temporal refinement ratio of 2.

At the outer boundary we impose Dirichlet conditions, with the metric
set to that of the initial data. These conditions are not strictly 
physically correct at finite radius, though the outer boundary is placed
sufficiently far that it is causally disconnected 
from the horizon for the time of the simulation.
We use black hole excision on the inner surface; namely, we 
find the apparent horizon (AH) using a flow method, and dynamically adjust
this boundary (the {\em excision surface}) to be some distance within the AH. Due to the causal nature of spacetime
inside the AH, no boundary conditions are placed on the excision surface. 

We adopt initial data describing a perturbed black string of mass per unit length $M$  
and length $L=20M\approx 1.4 L_c$ ($L_c$ is the critical length above which all perturbations
are unstable). This data was used in~\cite{Choptuik:2003qd} and
we refer the reader to that work for further details.

We evaluate the following curvature scalars on the AH:
\be\label{inv_def}
K=I R_{AH}^4/12, \ \ \ S=27 \left( 12 J^2 I^{-3}-1 \right ) + 1  \, ,
\ee
where $I=R_{abcd}R^{abcd}$, $J=R_{abcd} R^{cdef} R_{ef}{}^{ab}$ and  $R_{AH}$ is the
areal radius of the AH at the corresponding point (though note that
$I$ and $J$ are usually defined in terms of the Weyl tensor,
though here this equals the Riemann tensor as we are in vacuum).
$K$ and $S$ have been scaled
to evaluate to $\{6,1\}$ for the hyperspherical black hole and black
string respectively.

%
%

\noindent{\bf{\em Results:}} 
The results described here are from simulations where the computational domain
is $(r,z)\in ([0,320M]\times[0,20M])$. The coarsest grid covering the entire
domain has a resolution of $(N_r,N_z)=(1025,9)$ points. For convergence
studies we ran simulations with 3 values of the maximum estimated
truncation error $\tau$: [``low'',``medium'',``high''] resolution have
$\tau=[\tau_0,\tau_0/8,\tau_0/64]$ respectively.
This leads to an initial hierarchy where the horizon of the black string
is covered by 4, 5 and 6 additional refined levels for the 
low to high resolutions, respectively. Each
simulation was stopped when the estimated computational resources
required for continued evolution was prohibitively high (which naturally occurred
later in physical time for the lower resolutions); by then the
hierarchies were as deep as 17 levels.

\begin{figure}
\begin{center}
\includegraphics[width=3.in,clip=true]{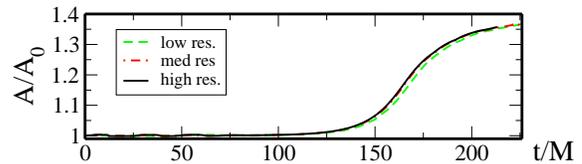}
\caption{(Normalized) apparent horizon area vs. time.
}
\label{fig:AH_radius}
\end{center}
\end{figure}

\begin{figure}
\begin{center}
\includegraphics[width=3.in,clip=true]{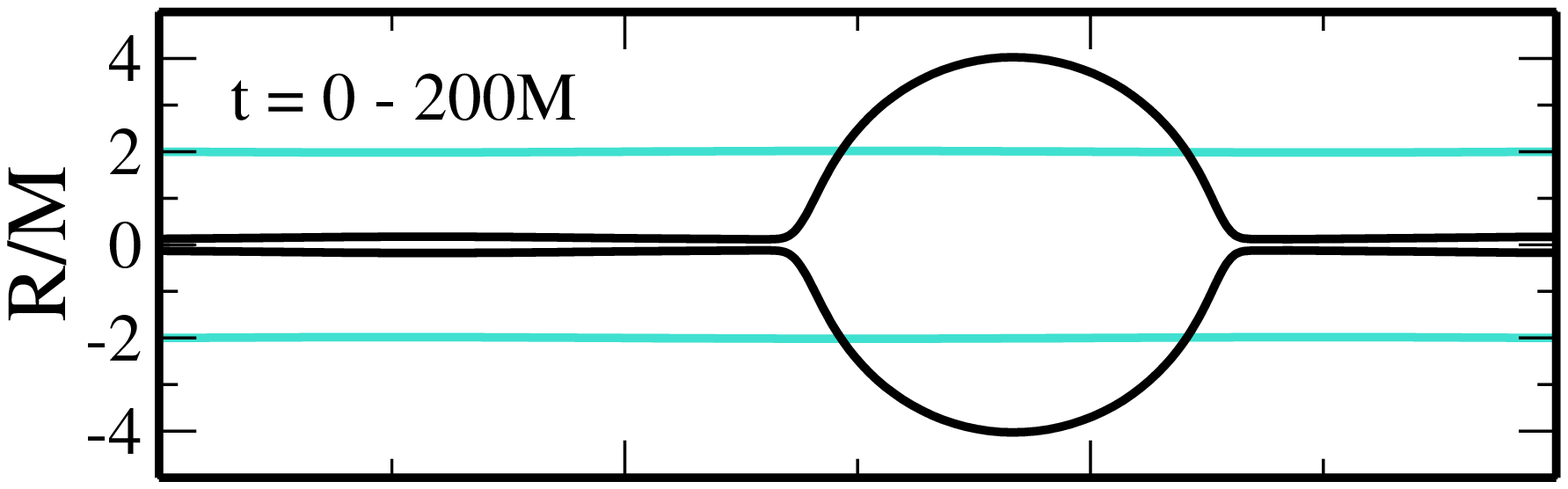}
\includegraphics[width=3.in,clip=true]{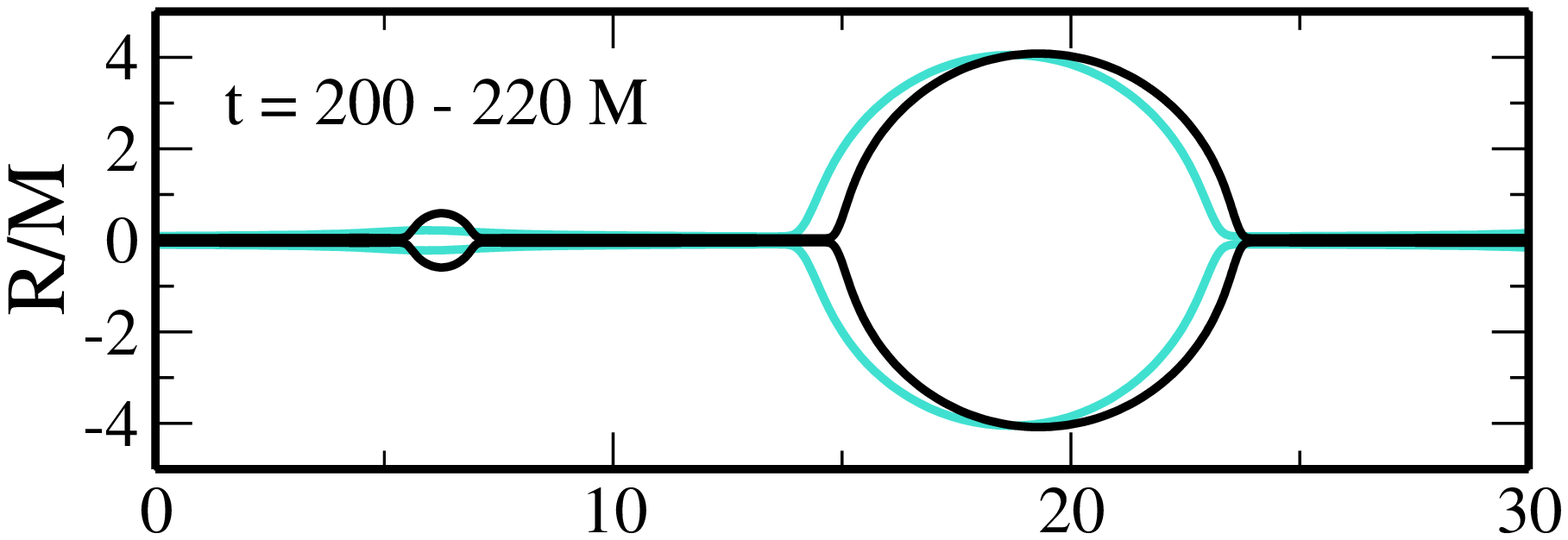}
\includegraphics[width=3.in,clip=true]{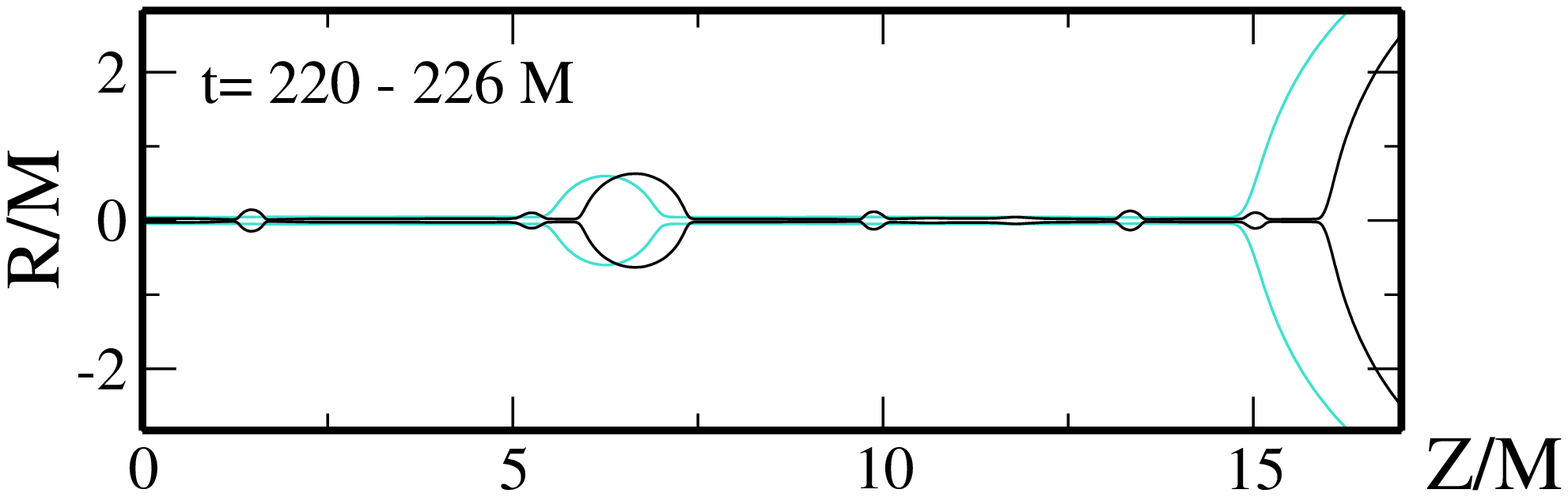}
\caption{Embedding diagram of the apparent horizon at several
instances in the evolution of the perturbed black string,
from the medium resolution run.
$R$ is areal radius, and the embedding coordinate
$Z$ is defined so that the proper length of the horizon in
the space-time $z$ direction (for a fixed $t,\theta,\phi$) is exactly equal to the 
Euclidean length of $R(Z)$ in the above figure. 
For visual aid copies of the diagrams reflected about $R=0$ have
also been drawn in.
The light (dark) lines denote the first (last) time from the time-segment
depicted in the corresponding panel.
The computational domain is periodic in $z$ with period $\delta z = 20M$; at  the
initial (final) time of the simulation  $\delta Z=20M$ ($\delta Z=27.2M$).  }
\label{fig:AH_embed}
\end{center}
\end{figure}

\begin{figure}
\begin{center}
\includegraphics[width=3.0in,clip=true]{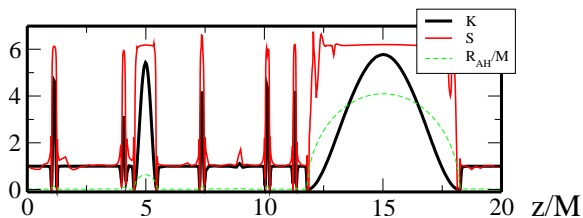}
\caption{Curvature invariants evaluated on the apparent horizon at the
last time of the simulation depicted in Fig.~\ref{fig:AH_embed}.
The invariant $K$ evaluates to $1$ for an exact black string, and $6$ for an exact
spherical black hole; similarly for  $S$ (\ref{inv_def}).
}
\label{fig:AH_invariants}
\end{center}
\end{figure}

Fig.~\ref{fig:AH_radius} shows the integrated AH area $A$ within $z\in[0,L]$
versus time. At the end of the lowest resolution run the total area
is $A=(1.369\pm0.005) A_0$\footnote{The error in the area was estimated
from convergence at the latest time data was available from all simulations}, where $A_0$ is the initial area; interestingly,
this almost reaches the value of $1.374 A_0$ that an exact 5D black hole
of the same total mass would have.
Fig.~\ref{fig:AH_embed} shows snapshots of embedding diagrams
of the AH, and Fig.~\ref{fig:AH_invariants} shows the curvature 
invariants (\ref{inv_def}) evaluated on the AH at the last time step,
both from the medium resolution run.

The shape of the AH, and that the invariants are 
tending to the limits associated with pure black strings or black
holes at corresponding locations on the AH, suggests it is  
reasonable to describe the local geometry as being similar
to a sequence of black holes connected by black strings.
This also strongly suggests that satellite formation will continue
self-similarly, as each string segment resembles
a uniform black string that is sufficiently long to be unstable.
Even if at some point in the cascade shorter segments were to form, 
this would not be a stable configuration as generically
the satellites will have some non-zero $z$-velocity, 
causing adjacent satellites to merge and effectively lengthening the 
connecting string segments.
With this interpretation, we summarize key features of the AH dynamics
in Table~\ref{tab_properties}. 

\begin{table}
{\small
\begin{tabular}[t]{| c || c | c | c | c | c |}
\hline
 Gen. & $t_i/M$ & $R_{s,i}/M$ & $L_{s,i}/R_{s,i}$ & $n_s$ & $R_{h,f}/M$\\
\hline
 1 & $118.1\pm0.5$ & $2.00$   & $10.0$ & $ 1 $ & $4.09\pm0.5\%$\\
\hline
 2 & $203.1\pm0.5$ & $0.148\pm1\%$ & $105\pm1\%$ & $ 1 $ & $0.63\pm2\%$\\
\hline
 3 & $223\pm2$     & $0.05\pm20\%$ & $\approx  10^2$ & $ >1$ & $0.1 - 0.2$ \\
\hline
 4 & $\approx 227$ & $\approx 0.02 $ & $\approx  10^2$ & $ >1(?)$ & ?      \\
\hline
\end{tabular}
}
\caption{Properties of the evolving black string apparent horizon, 
{\em interpreted} as proceeding through several
self-similar generations, where each local string segment temporarily 
reaches a near-steady state before the onset of the next GL instability.
$t_i$ is the time when the instability has grown to where
the nascent spherical
region reaches an areal radius $1.5$ times
the surrounding string-segment radius $R_{s,i}$, which has
an estimated proper length $L_{s,i}$ (the critical $L/R$ is $\approx 7.2$~\cite{Gregory:1993vy}). 
$n_s$ is the number of satellites that form per segment, that
each attain a radius $R_{h,f}$ measured at the end of the simulation. 
Errors, where appropriate, come from convergence tests.
After the second generation
the number and distribution of satellites that form
depend sensitively on grid parameters, and perhaps the only
``convergent'' result we have then 
is at roughly $t=223$ a third generation {\em does} develop. 
We surmise the reason for this is the long
parent string segments could have multiple unstable modes
with similar growth rates, and which is first excited 
is significantly affected by truncation error.
We have only had the resources to run the lowest resolution simulation for sufficiently long 
to see the onset of the 4th generation, hence the lack of error estimates
and presence of question marks in the corresponding row.
}
\label{tab_properties}
\end{table}

\begin{figure}
\begin{center}
\includegraphics[width=3.2in,clip=true]{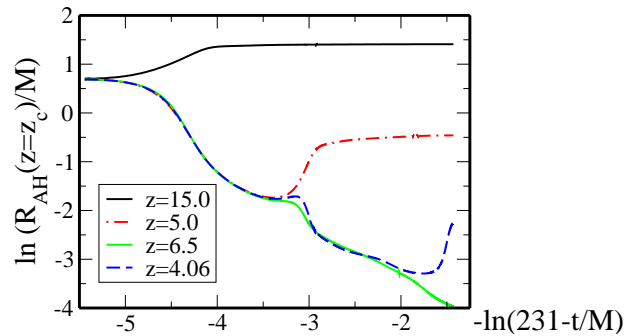}
\caption{
Logarithm of the areal  radius vs. logarithm of time
for select points on the apparent horizon from
the simulation depicted in Fig.~\ref{fig:AH_embed}. 
We have shifted the time axis {\em assuming} self-similar behavior;
the putative naked singularity forms at asymptotic time $t/M\approx 231$.
The coordinates
at $z=15,5$ and $4.06$ correspond to the maxima of the areal
radii of the first and second generation satellites, and
one of the third generation satellites at the time the simulation
stopped.
The value $z=6.5$ is a representative slice in the middle of a
piece of the horizon that remains string-like throughout the evolution.
}
\label{fig:AH_radius_vs_lnt}
\end{center}
\end{figure}

We estimate when this self-similar cascade will end.
The time when the first satellite appears is controlled
by the perturbation imparted by the initial data; here that
is $T_0/M\approx 118$. 
Subsequent timescales should approximately represent the generic 
development of the instability. The time for the 
first instability {\em after} that sourced by the initial data 
is $T_1/M\approx 80$. Beyond that, with the caveats that we have
a small number of points and poor control over errors at late
times, each subsequent instability unfolds
on a time-scale $X\approx1/4$ times that of the preceding one.
This is to be expected if, as for the exact black string,
%
the time scale is proportional to the string radius. The time $t_0$
of the end-state is then $t_0 \approx T_0 + \sum_{i=0}^\infty T_1 X^i = T_0 + T_1/(1-X)$.
For the data here, $t_0 /M\approx 231$; then the local
string segments reach zero radius, and 
the curvature visible to exterior observers diverges.
Fig.~\ref{fig:AH_radius_vs_lnt} shows a few points on the AH, scaled 
assuming this behavior. In the Rayleigh-Plateau analogue,
the shrinking neck of a fluid stream has a self-similar scaling solution
that satisfies $r\propto (t_0-t)$, or, $d\ln r/d(-\ln(t_0-t))=-1$, 
where $r$ is the stream radius (see ~\cite{eggers_pinch}, and~\cite{miyamoto_extend}
for extensions to higher dimensions);
to within $10-20\%$ this is the average slope we see (e.g. Fig.~\ref{fig:AH_radius_vs_lnt})
at string segments of the AH at late times.

%
%
\noindent{\bf{\em Conclusions:}} 
We have
studied the dynamics of a perturbed, 
unstable 5D black string. The 
horizon behaves 
similarly to 
 the surface of a stream of low viscosity
fluid subject to the Rayleigh-Plateau instability. Multiple
generations of spherical satellites, connected by ever thinner string
segments, form. Curvature invariants on the horizon suggest
this is a self-similar process, where at each stage the local string/spherical
segments resemble the corresponding exact solutions. Furthermore, the time scale
for the formation of the next generation is proportional
to the local string radius, implying the cascade will terminate in finite
asymptotic time. Since local curvature scalars grow with inverse
powers of the string radius, this end-state will thus be a naked, curvature singularity.
If quantum gravity resolves these singularities, a series of spherical
black holes will emerge. However, small
momentum perturbations in the extra dimension would induce the merger of 
these black holes, thus for a compact
extra dimension the end state of the GL instability will be a single
black hole with spherical topology.

The kind of singularity reached here via a self-similar process is
akin to that formed in critical gravitational collapse~\cite{Choptuik:1992jv}; however,
here no fine-tuning is required. Thus,
5 (and presumably higher) dimensional Einstein gravity allows solutions
that generically violate cosmic censorship. 
Angular momentum will likely not alter this conclusion, since
as argued in~\cite{Emparan:2009at}, and shown
in~\cite{Dias:2010eu}, rotation does not suppress 
the unstable modes, and moreover induces super-radiant
and gyrating instabilities ~\cite{Marolf:2004fya}.



%
%
\noindent{\bf{\em Acknowledgments:}}
We thank  
V. Cardoso, M. Choptuik, R. Emparan, D. Garfinkle, K. Lake, 
S. Gubser, G. Horowitz, D. Marolf, R. Myers,
W. Unruh and R. Wald for stimulating discussions.
This work was supported by NSERC (LL), CIFAR (LL),
the Alfred P. Sloan Foundation (FP), and NSF grant PHY-0745779 (FP).
Simulations were run on the {\bf Woodhen} cluster at Princeton University and
at LONI. Research at Perimeter Institute is
supported through Industry Canada and by the Province of Ontario
through the Ministry of Research \& Innovation.
%
%
\bibliographystyle{apsrev}

\begin{thebibliography}{25}
\expandafter\ifx\csname natexlab\endcsname\relax\def\natexlab#1{#1}\fi
\expandafter\ifx\csname bibnamefont\endcsname\relax
  \def\bibnamefont#1{#1}\fi
\expandafter\ifx\csname bibfnamefont\endcsname\relax
  \def\bibfnamefont#1{#1}\fi
\expandafter\ifx\csname citenamefont\endcsname\relax
  \def\citenamefont#1{#1}\fi
\expandafter\ifx\csname url\endcsname\relax
  \def\url#1{\texttt{#1}}\fi
\expandafter\ifx\csname urlprefix\endcsname\relax\def\urlprefix{URL }\fi
\providecommand{\bibinfo}[2]{#2}
\providecommand{\eprint}[2][]{\url{#2}}

\bibitem[{\citenamefont{Gregory and Laflamme}(1993)}]{Gregory:1993vy}
\bibinfo{author}{\bibfnamefont{R.}~\bibnamefont{Gregory}} \bibnamefont{and}
  \bibinfo{author}{\bibfnamefont{R.}~\bibnamefont{Laflamme}},
  \bibinfo{journal}{Phys. Rev. Lett.} \textbf{\bibinfo{volume}{70}},
  \bibinfo{pages}{2837} (\bibinfo{year}{1993}).

\bibitem[{\citenamefont{Emparan and Reall}(2008)}]{Emparan:2008eg}
\bibinfo{author}{\bibfnamefont{R.}~\bibnamefont{Emparan}} \bibnamefont{and}
  \bibinfo{author}{\bibfnamefont{H.~S.} \bibnamefont{Reall}},
  \bibinfo{journal}{Living Rev. Rel.} \textbf{\bibinfo{volume}{11}},
  \bibinfo{pages}{6} (\bibinfo{year}{2008});
\bibinfo{author}{\bibfnamefont{T.}~\bibnamefont{Harmark}},
  \bibinfo{author}{\bibfnamefont{V.}~\bibnamefont{Niarchos}}, \bibnamefont{and}
  \bibinfo{author}{\bibfnamefont{N.~A.} \bibnamefont{Obers}},
  \bibinfo{journal}{Class. Quant. Grav.} \textbf{\bibinfo{volume}{24}},
  \bibinfo{pages}{R1} (\bibinfo{year}{2007});
\bibinfo{author}{\bibfnamefont{B.}~\bibnamefont{Kol}}, \bibinfo{journal}{Phys.
  Rept.} \textbf{\bibinfo{volume}{422}}, \bibinfo{pages}{119}
  (\bibinfo{year}{2006}).

\bibitem[{\citenamefont{{Hawking} and {Ellis}}(1975)}]{Hawking:1973uf}
\bibinfo{author}{\bibfnamefont{S.~W.} \bibnamefont{{Hawking}}}
  \bibnamefont{and} \bibinfo{author}{\bibfnamefont{G.~F.~R.}
  \bibnamefont{{Ellis}}}, \emph{\bibinfo{title}{{The Large Scale Structure of
  Space-Time}}} (\bibinfo{publisher}{Cambridge University Press},
  \bibinfo{year}{1975}).

\bibitem[{\citenamefont{Horowitz and Maeda}(2001)}]{Horowitz:2001cz}
\bibinfo{author}{\bibfnamefont{G.~T.} \bibnamefont{Horowitz}} \bibnamefont{and}
  \bibinfo{author}{\bibfnamefont{K.}~\bibnamefont{Maeda}},
  \bibinfo{journal}{Phys. Rev. Lett.} \textbf{\bibinfo{volume}{87}},
  \bibinfo{pages}{131301} (\bibinfo{year}{2001}).

\bibitem[{\citenamefont{Wiseman}(2003)}]{Wiseman:2002zc}
\bibinfo{author}{\bibfnamefont{T.}~\bibnamefont{Wiseman}},
  \bibinfo{journal}{Class. Quant. Grav.} \textbf{\bibinfo{volume}{20}},
  \bibinfo{pages}{1137} (\bibinfo{year}{2003}).

\bibitem[{\citenamefont{Gubser}(2002)}]{Gubser:2001ac}
\bibinfo{author}{\bibfnamefont{S.~S.} \bibnamefont{Gubser}},
  \bibinfo{journal}{Class. Quant. Grav.} \textbf{\bibinfo{volume}{19}},
  \bibinfo{pages}{4825} (\bibinfo{year}{2002}).

\bibitem[{\citenamefont{Sorkin}(2004)}]{Sorkin:2004qq}
\bibinfo{author}{\bibfnamefont{E.}~\bibnamefont{Sorkin}},
  \bibinfo{journal}{Phys. Rev. Lett.} \textbf{\bibinfo{volume}{93}},
  \bibinfo{pages}{031601} (\bibinfo{year}{2004}).

\bibitem[{\citenamefont{Choptuik et~al.}(2003)}]{Choptuik:2003qd}
\bibinfo{author}{\bibfnamefont{M.~W.} \bibnamefont{Choptuik}}
  \bibnamefont{et~al.}, \bibinfo{journal}{Phys. Rev.}
  \textbf{\bibinfo{volume}{D68}}, \bibinfo{pages}{044001}
  (\bibinfo{year}{2003}).

\bibitem[{\citenamefont{Garfinkle et~al.}(2005)\citenamefont{Garfinkle, Lehner,
  and Pretorius}}]{Garfinkle:2004em}
\bibinfo{author}{\bibfnamefont{D.}~\bibnamefont{Garfinkle}},
  \bibinfo{author}{\bibfnamefont{L.}~\bibnamefont{Lehner}}, \bibnamefont{and}
  \bibinfo{author}{\bibfnamefont{F.}~\bibnamefont{Pretorius}},
  \bibinfo{journal}{Phys. Rev.} \textbf{\bibinfo{volume}{D71}},
  \bibinfo{pages}{064009} (\bibinfo{year}{2005}).

\bibitem[{\citenamefont{Marolf}(2005)}]{Marolf:2005vn}
\bibinfo{author}{\bibfnamefont{D.}~\bibnamefont{Marolf}},
  \bibinfo{journal}{Phys. Rev.} \textbf{\bibinfo{volume}{D71}},
  \bibinfo{pages}{127504} (\bibinfo{year}{2005}).

\bibitem{Thorne:1986iy}
  K.~S.~.~Thorne, R.~H.~.~Price and D.~A.~.~Macdonald,
  ``Black holes: The Membrane Paradigm,''
  {\it  New Haven, USA: Yale Univ. Pr. (1986) 367p}

\bibitem[{\citenamefont{Bhattacharyya et~al.}(2008)\citenamefont{Bhattacharyya,
  Hubeny, Minwalla, and Rangamani}}]{Bhattacharyya:2008jc}
\bibinfo{author}{\bibfnamefont{S.}~\bibnamefont{Bhattacharyya}},
  \bibinfo{author}{\bibfnamefont{V.~E.} \bibnamefont{Hubeny}},
  \bibinfo{author}{\bibfnamefont{S.}~\bibnamefont{Minwalla}}, \bibnamefont{and}
  \bibinfo{author}{\bibfnamefont{M.}~\bibnamefont{Rangamani}},
  \bibinfo{journal}{JHEP} \textbf{\bibinfo{volume}{02}}, \bibinfo{pages}{045}
  (\bibinfo{year}{2008}).

\bibitem[{\citenamefont{Emparan et~al.}(2010)\citenamefont{Emparan, Harmark,
  Niarchos, and Obers}}]{Emparan:2009at}
\bibinfo{author}{\bibfnamefont{R.}~\bibnamefont{Emparan}},
  \bibinfo{author}{\bibfnamefont{T.}~\bibnamefont{Harmark}},
  \bibinfo{author}{\bibfnamefont{V.}~\bibnamefont{Niarchos}}, \bibnamefont{and}
  \bibinfo{author}{\bibfnamefont{N.~A.} \bibnamefont{Obers}},
  \bibinfo{journal}{JHEP} \textbf{\bibinfo{volume}{03}}, \bibinfo{pages}{063}
  (\bibinfo{year}{2010}).

\bibitem[{\citenamefont{Cardoso and Dias}(2006)}]{Cardoso:2006ks}
\bibinfo{author}{\bibfnamefont{V.}~\bibnamefont{Cardoso}} \bibnamefont{and}
  \bibinfo{author}{\bibfnamefont{O.~J.~C.} \bibnamefont{Dias}},
  \bibinfo{journal}{Phys. Rev. Lett.} \textbf{\bibinfo{volume}{96}},
  \bibinfo{pages}{181601} (\bibinfo{year}{2006}).

\bibitem[{\citenamefont{Cardoso and Gualtieri}(2006)}]{Cardoso:2006sj}
\bibinfo{author}{\bibfnamefont{V.}~\bibnamefont{Cardoso}} \bibnamefont{and}
  \bibinfo{author}{\bibfnamefont{L.}~\bibnamefont{Gualtieri}},
  \bibinfo{journal}{Class. Quant. Grav.} \textbf{\bibinfo{volume}{23}},
  \bibinfo{pages}{7151} (\bibinfo{year}{2006}).

\bibitem{Camps:2010br}
  J.~Camps, R.~Emparan and N.~Haddad,
  JHEP {\bf 1005}, 042 (2010)

\bibitem[{\citenamefont{Eggers}(1997)}]{Eggers:1997zz}
\bibinfo{author}{\bibfnamefont{J.}~\bibnamefont{Eggers}},
  \bibinfo{journal}{Rev. Mod. Phys.} \textbf{\bibinfo{volume}{69}},
  \bibinfo{pages}{865} (\bibinfo{year}{1997}).

\bibitem[{\citenamefont{Kovtun et~al.}(2005)\citenamefont{Kovtun, Son, and
  Starinets}}]{Kovtun:2004de}
\bibinfo{author}{\bibfnamefont{P.}~\bibnamefont{Kovtun}},
  \bibinfo{author}{\bibfnamefont{D.~T.} \bibnamefont{Son}}, \bibnamefont{and}
  \bibinfo{author}{\bibfnamefont{A.~O.} \bibnamefont{Starinets}},
  \bibinfo{journal}{Phys. Rev. Lett.} \textbf{\bibinfo{volume}{94}},
  \bibinfo{pages}{111601} (\bibinfo{year}{2005}).

\bibitem[{\citenamefont{Pretorius}(2005)}]{Pretorius:2004jg}
\bibinfo{author}{\bibfnamefont{F.}~\bibnamefont{Pretorius}},
  \bibinfo{journal}{Class. Quant. Grav.} \textbf{\bibinfo{volume}{22}},
  \bibinfo{pages}{425} (\bibinfo{year}{2005}).

\bibitem[{\citenamefont{Alcubierre et~al.}(2001)}]{Alcubierre:1999ab}
\bibinfo{author}{\bibfnamefont{M.}~\bibnamefont{Alcubierre}}
  \bibnamefont{et~al.}, \bibinfo{journal}{Int. J. Mod. Phys.}
  \textbf{\bibinfo{volume}{D10}}, \bibinfo{pages}{273} (\bibinfo{year}{2001}).

\bibitem[{\citenamefont{Gundlach et~al.}(2005)\citenamefont{Gundlach,
  Martin-Garcia, Calabrese, and Hinder}}]{Gundlach:2005eh}
\bibinfo{author}{\bibfnamefont{C.}~\bibnamefont{Gundlach}},
  \bibinfo{author}{\bibfnamefont{J.~M.} \bibnamefont{Martin-Garcia}},
  \bibinfo{author}{\bibfnamefont{G.}~\bibnamefont{Calabrese}},
  \bibnamefont{and} \bibinfo{author}{\bibfnamefont{I.}~\bibnamefont{Hinder}},
  \bibinfo{journal}{Class. Quant. Grav.} \textbf{\bibinfo{volume}{22}},
  \bibinfo{pages}{3767} (\bibinfo{year}{2005}).

\bibitem{eggers_pinch} J.Eggers, Phys.Rev.Lett {\bf 71}, 3458 (1993).

\bibitem{miyamoto_extend} U.~Miyamoto, 
  arXiv:1007.4302 [hep-th];
  U.~Miyamoto, {\em to be published}

\bibitem[{\citenamefont{Choptuik}(1993)}]{Choptuik:1992jv}
\bibinfo{author}{\bibfnamefont{M.~W.} \bibnamefont{Choptuik}},
  \bibinfo{journal}{Phys. Rev. Lett.} \textbf{\bibinfo{volume}{70}},
  \bibinfo{pages}{9} (\bibinfo{year}{1993}).

\bibitem{Dias:2010eu}
  O.~J.~C.~Dias, P.~Figueras, R.~Monteiro, H.~S.~Reall and J.~E.~Santos,
  JHEP {\bf 1005}, 076 (2010);

\bibitem{Marolf:2004fya}
  D.~Marolf and B.~C.~Palmer,
  Phys.\ Rev.\  D {\bf 70}, 084045 (2004);
\bibinfo{author}{\bibfnamefont{V.}~\bibnamefont{Cardoso}} \bibnamefont{and}
  \bibinfo{author}{\bibfnamefont{J.~P.~S.} \bibnamefont{Lemos}},
  \bibinfo{journal}{Phys. Lett.} \textbf{\bibinfo{volume}{B621}},
  \bibinfo{pages}{219} (\bibinfo{year}{2005}).
  V.~Cardoso and S.~Yoshida,
  JHEP {\bf 0507}, 009 (2005)


\end{thebibliography}

%
%
\end{document}